\documentclass[reprint,amsmath,amssymb,aps,prl,superscriptaddress,]{revtex4-2}

\usepackage{subfiles}
\usepackage{graphicx}
\usepackage{dcolumn}
\usepackage{bm}
\usepackage{color}
\usepackage{braket}
\usepackage{hyperref}
\hypersetup{hidelinks}

\usepackage{graphicx}
\usepackage{dcolumn}
\usepackage{bm}
\usepackage{color}
\usepackage{textcomp}

\begin{document}

\preprint{APS/123-QED}

\title{Communicating at a record 14.5 bits per received photon through a photon-starved channel}

\author{Sai Kanth Dacha}
\thanks{These authors contributed equally to this work.}
\affiliation{Nokia Bell Labs, Murray Hill, 600 Mountain Ave, New Providence, NJ 07974, United States}
\affiliation{Department of Applied Physics and Applied Mathematics, Columbia University, New York, NY 10027, United States}
\author{Ren\'e-Jean Essiambre}
\thanks{These authors contributed equally to this work.}
\affiliation{Nokia Bell Labs, Murray Hill, 600 Mountain Ave, New Providence, NJ 07974, United States}
\author{Alexei Ashikhimin}
\affiliation{Nokia Bell Labs, Murray Hill, 600 Mountain Ave, New Providence, NJ 07974, United States}
\author{Andrea Blanco-Redondo}
\affiliation{Nokia Bell Labs, Murray Hill, 600 Mountain Ave, New Providence, NJ 07974, United States}
\author{Frank R. Kschischang}
\affiliation{Department of Electrical and Computer Engineering, University of Toronto, Toronto, ON M5S 3G4, Canada}
\author{Konrad Banaszek}
\affiliation{Centre for Quantum Optical Technologies, University of Warsaw, 02-097 Warszawa, Poland}
\author{Yuanhang Zhang}
\affiliation{Nokia Bell Labs, Murray Hill, 600 Mountain Ave, New Providence, NJ 07974, United States}

\date{22 January 2025}%

\begin{abstract}
Exploration of the Universe requires communication with Earth, either on a direct path or through a cascade of proximate celestial bodies. Microwaves have traditionally been used for space communication, but electromagnetic waves of higher frequencies, such as in the optical domain, will enable probing farther in space due to their considerably lower diffraction loss. At a given data rate, the ultimate limit to point-to-point optical communication is determined by the received signal power and the photon information efficiency. The latter measures the number of information bits extracted per photon incident on a detector. As distances across space and, consequently, path loss increases, the system that can achieve the highest photon information efficiency will determine the longest distance at which communication is possible. We report here an experimental demonstration of optical detection at a record photon information efficiency of 14.5 bits per incident photon, or 17.8 bits per detected photon, after 87.5~dB of attenuation. Expressed in terms of energy per bit, this corresponds to 8.84~zeptojoules per bit, or 0.069~photons per bit at 1550~nm. To our knowledge, this is the highest photon information efficiency or lowest energy per bit detection system ever demonstrated at optical frequencies. Such a sensitive detection system holds promise for a wide range of applications. 
\end{abstract}

\maketitle


\section{Introduction}
\label{sec:intro}  
The ability to detect information using the lowest possible incident power on a receiver drives several fundamental applications, such as covert communication~\cite{ataie2015subnoise}, green networking~\cite{tucker2010green}, and transmission of information through a high-attenuation channel~\cite{caplan2007laser}. 
In this regime of low energy per bit at detection, it is convenient to use a metric known as Photon Information Efficiency (PIE), expressed in the number of bits conveyed by each photon. 

\begin{figure*}[htbp]
\centering
\includegraphics[width=\linewidth]{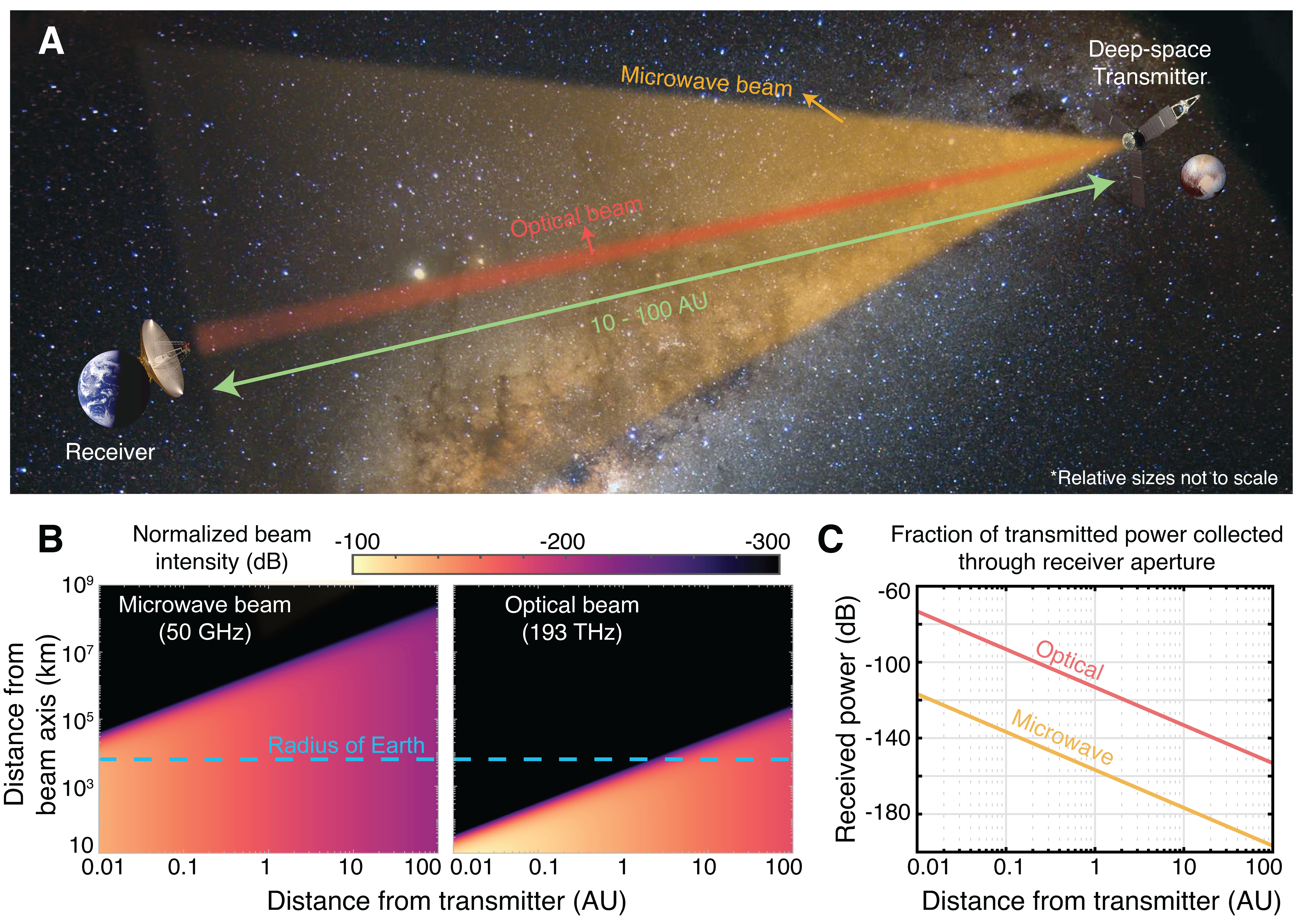}
\caption{Comparison of microwave and optical carrier beams for downlink deep-space communication. (a) Illustration depicting a deep-space link connecting a distant spacecraft with an Earth-based receiver. Electromagnetic waves with microwave carrier frequencies experience significantly larger diffraction compared to optical carriers. (b) Colormap plot, on logarithmic scales, describing the intensity of the microwave (50 GHz) and optical (193 THz) beams, normalized to their respective on-axis maxima at the beam waist. For distances larger than a few astronomical units (AU), the transversal extents of the received beams are larger than the radius of Earth (blue dashed line), with the microwave beam being over three orders of magnitude larger than the optical beam for the same propagation distance. (c) Fraction of transmitted power collected through receiver aperture in the far-field, calculated for microwave and optical beams, for transmitter and receiver aperture diameters ($D_R$, $D_T$) of (3.7 m, 20 m) and (0.2 m, 2.28 m), respectively. The power loss experienced by the microwave beam is several orders of magnitude larger. [Images from NASA/ESA]}
\label{fig:fig0}
\end{figure*}

Deep-space communication may represent the most extreme high-attenuation channel. Communication between deep-space probes and receivers near Earth uses electromagnetic waves with carriers in the microwave range, typically ranging from a few GHz (X band) to a few tens of GHz (Ka band)~\cite{shin2015frequency}, with coherent detection at the receiver~\cite{caplan2007laser}. However, long-distance microwave-based communication suffers from much greater diffraction loss~\cite{friis1946note} than optical waves~\cite{kogelnik1966laser,jarzyna2024photon} because the beam size is much larger due to the longer wavelengths involved, as shown in Figures~\ref{fig:fig0}(a) and \ref{fig:fig0}(b).

In the far-field approximation, the signal power incident on a receiver is given by $P_R(d) = P_T \, (\pi D_T D_R  f_c/ (4cd))^2$, where $P_T$ is the transmit power, $f_c$ the carrier frequency, $d$ the propagation distance, and $D_T$ and $D_R$ the transmit and receive aperture diameters, respectively. At a given transmit power and fixed aperture sizes, the received signal power $P_R(d)$ impinging on a distant detector scales as $f_c^2$. Figure~\ref{fig:fig0}(c) shows a comparison of the net power loss for microwave and optical carrier frequencies, for typical transmit and receive apertures. On the other hand, the quantum noise power, $P_n$, scales as $P_n \propto h f_c$, where $h$ is the Planck constant. As a result, the signal-to-noise ratio (SNR) scales linearly with the carrier frequency, that is ${\rm SNR}(d) = P_R(d) / P_n \propto f_c$. This provides a strong incentive to move from microwaves to optical frequencies for deep-space communication, since optical carriers operate at frequencies around 200~THz, leading to a potential SNR improvement by four orders of magnitude. In addition, operating at optical frequencies allows spatial beam confinement to improve privacy and minimize interference (or jamming), as well as providing orders of magnitude larger and mostly unregulated bandwidth.

Over the past few decades, tremendous progress has been made in increasing the PIE of optical communication links~\cite{essiambre2010capacity,burenkov2021practical}, with a key technology being low-noise high-quantum-efficiency photodetectors, such as superconducting nanowire single-photon detectors (SNSPDs)~\cite{you2020superconducting}. The first technology demonstration of deep-space optical communication using quantum detectors on Earth is the Psyche Mission~\cite{nasa2023psyche} launched in October 2023.

In his landmark 1948 paper~\cite{sha1948}, Claude Shannon presented a general method to calculate the maximum rate of transmission of information, referred to as the channel capacity $C$, over a noisy communication channel. Shannon explicitly considered both signal and noise fields as continuous waves with the ability of the receiver to detect full fields. Using Shannon's information theory formulation, and taking into account the quantum nature of electromagnetic waves, James Gordon predicted in 1962~\cite{gordon1962quantum} that capacity at detection can be further increased using photon counters. The optimum information rate was later found by Holevo and co-workers~\cite{holevo1998capacity,Giovannetti2014} to be the upper bound of capacity for arbitrary prepared quantum states. This limit is referred to as the Gordon-Holevo capacity limit. 
The PIE limit of a communication system is obtained by dividing the Gordon-Holevo capacity by the rate of arrival of photons at the receiver. For a given transmit power, channel attenuation, and data rate, only systems achieving a sufficiently high PIE can usefully transmit information. 

\begin{figure}[htbp]
\centering
\includegraphics[width=\linewidth]{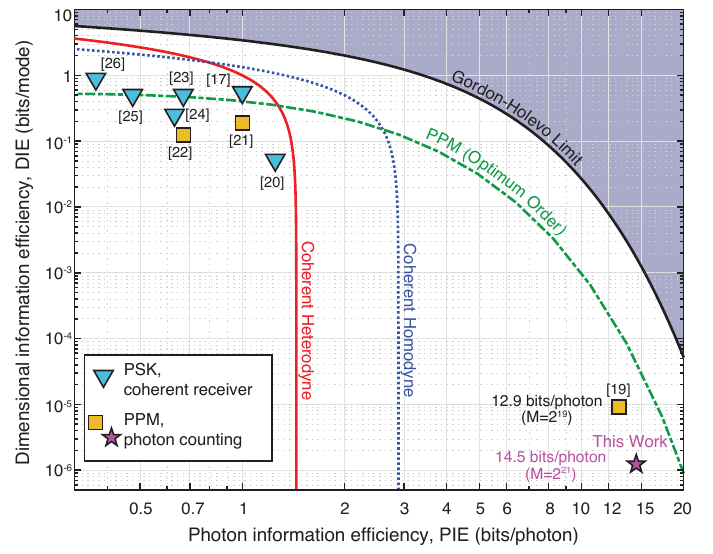}
\caption{Theoretical limits and experimentally demonstrated photon information efficiency (PIE) and their associated dimensional information efficiency (DIE). The various curves  depict theoretical capacities for photon counters (Gordon-Holevo), coherent homodyne and heterodyne, as well as for PPM modulation with optimum modulation order and photon counters. The symbols display key experimental demonstrations cited in the references section of this paper. The highest PIE enables communication with the faintest received signal. (M: order of Pulse-Position Pulse (PPM), PSK: Phase-Shift Keying)}
\label{fig:DIEvsPIE}
\end{figure}

The theoretical PIE limits for the most prominent optical detection technologies~\cite{dolinar2012ultimate,kakarla2020one,banaszek2020quantum}, and several key experimental demonstrations~\cite{Farr2013,E2,kakarla2020one,E4,E5,E6,E7,E8,E9}, are shown in Fig.~\ref{fig:DIEvsPIE}. The dimensional information efficiency (DIE) is a measure of the number of bits transmitted per ``mode''. In deep-space exploration, the receiver aperture is much smaller than the beam size, favoring using time, i.e. time slots, rather than space, as transmission modes to increase power efficiency. The PIE is calculated from the channel capacity $C$ using PIE = $C/F_s$, where $F_s = P_s/(h f_c)$ is the rate of arrival of signal photons, where $P_s$ is the signal power and $h f_c$ the energy of a single photon. The DIE is calculated using DIE = $C/R_\mathrm{slot}$, where $R_\mathrm{slot}$ is the time slots arrival rate simply given by $R_\mathrm{slot} = 1 / \mathrm{T_{slot}}$, where $\mathrm{T_{slot}}$ is the time slot duration.

For systems limited solely by quantum noise, that is, in the absence of excess noise, the Gordon-Holevo quantum capacity limit $C_\mathrm{GH}$ per unit of bandwidth $B$ is given by~\cite{gordon1962quantum,holevo1998capacity,Giovannetti2014} $C_\mathrm{GH}/B = \log_2(1+n_s) + n_s \, \log_2(1+1/n_s)$, where $n_s = P_s/(h f_c B) = F_s/B$, or the rate of arrival of photons per unit of signal bandwidth $B$~\cite{moision2014range}. Fig.~\ref{fig:DIEvsPIE} shows the PIE limit associated with the Gordon-Holevo capacity limit. It represents the ultimate limit in the power efficiency of communication. Also displayed are the PIE limits of two forms of coherent detection, the conventional heterodyne detection and the ultimate homodyne~\cite{banaszek2020quantum}. The capacity of ideal heterodyne detection $C_{\rm e}$ is given by $C_{\rm e}/B = \log_2(1 + n_s)$, while the limit for ideal homodyne detection $C_{\rm o}$ is given by $C_{\rm o}/B = 1/2 \, \log_2(1 + 4 n_s)$. Coherent heterodyne detection reaches a DIE higher than that of homodyne detection at low PIE, while it is the opposite at high PIE.
The heterodyne and homodyne PIE limits shown in Fig.~\ref{fig:DIEvsPIE} also correspond to the limits of an ideal phase-insensitive amplifier (PIA) and phase-sensitive amplification (PSA), respectively~\cite{tong2011towards,andrekson2020fiber,banaszek2020quantum}.

A practical modulation technique to achieve high PIE using time slots, also called time-bin encoding, is the Pulse-Position Modulation (PPM) format~\cite{jarzyna2024photon}, where one pulse is positioned in a temporal frame composed of $M=2^m$ time slots, where $m$ is an integer and $M$ is referred to as the PPM order. The expression for the PPM capacity at optimum order $M$ is given by $C_\mathrm{PPM}^\mathrm{opt} = (1 - \exp(-\lambda))^2 / ( (\lambda \log(2))(1 - (1+\lambda) \exp(-\lambda)) )$, where $\lambda = F_s \times \mathrm{T_{slot}}$ is the mean number of photons per PPM frame incident on the detector~\cite{Dolinar2011}. The optimal PPM order PIE curve is displayed in Fig.~\ref{fig:DIEvsPIE}, from which one can infer that, in the absence of excess noise, a large value of $M$ is necessary to achieve a high PIE. An important challenge in experimentally realizing high PIE using a large $M$ is the stringent requirement on the extinction ratio (here $\sim 80$~dB and higher) of the modulated laser, defined as the ratio of power in the time slot containing the data pulse to the average power per ``empty'' time slot. 

Additionally, to maximize the data rate, one should minimize the time-slot duration (less than $\sim 1$~ns). Developing an optical source that generates on-demand short pulses with a very high extinction ratio is an important challenge, especially when a low-complexity transmitter design is required, such as for downlink communication in deep-space exploration. 


In this Article, we report the experimental demonstration of a record PIE of 14.5 bits/photon using a large PPM order, $M=2^{21}$. This is enabled using a highly compact transmitter, that is powered by a single 5 Volts USB cable connected to a laptop, that directly modulates a distributed feedback (DFB) laser to generate very high average extinction ratio ($>90$~dB) short ($\mathrm{T_{pulse}}\approx50$~ps) optical pulses on demand.

\section{Experiment}
 \label{sec:experiment}
\begin{figure}[htbp]
\centering
\includegraphics[width=\linewidth]{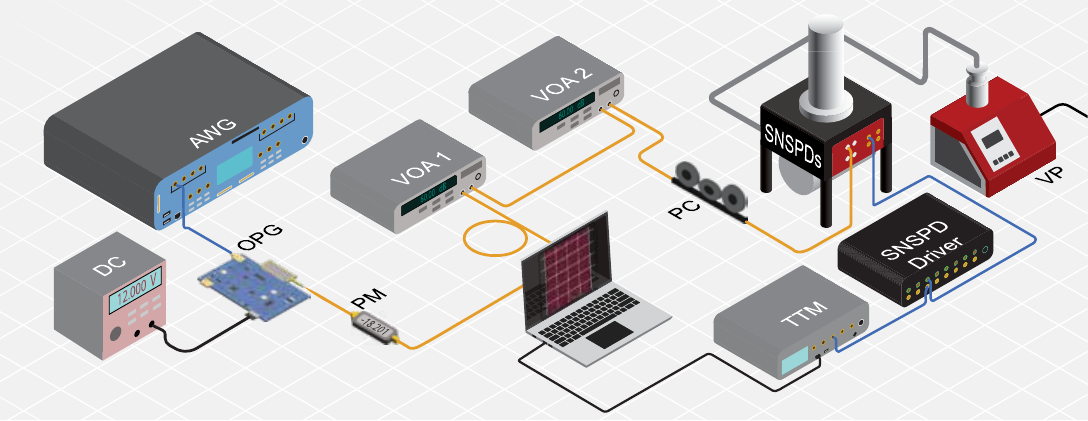}
\caption{Experimental schematic of communication testbed using single-photon detectors. AWG: arbitrary waveform generator, OPG: optical pulse generator, DC: direct current (power supply), PM: (inline) power meter, VOA: variable optical attenuator, PC: polarization controller, SNSPD: superconducting nanowire single-photon detector, VP: vacuum pump, TTM: time-tagging module.}
\label{fig:expschematic}
\end{figure}

Fig.~\ref{fig:expschematic} illustrates the experimental schematic. A computer-generated random bits sequence is segmented into bit strings of lengths $m=\mathrm{log}_2(M)$, each of which is mapped to a unique discrete temporal position of a pulse inside a PPM frame consisting of $M$ time slots. The temporal width of each time slot is $\mathrm{T_{slot}}\approx400$~ps. An empty guard window of $\approx100$~ns (larger than the SNSPD's dead time of $\approx60$~ns, and much shorter than the PPM frame duration $\sim0.1-1$~ms) is inserted between consecutive PPM frames, resulting in a ``guarded PPM frame" whose temporal duration is given as $\mathrm{T_{frame}}=M\mathrm{T_{slot}}+100$~ns.

The resulting digital signal is loaded onto a programmable arbitrary waveform generator (AWG) with a large dynamic memory (Keysight M8190A), which triggers an optical pulse generator circuit (OPG; Highland Technologies T165 laser diode pulser) that directly modulates a DFB laser to produce optical pulses with a full-width at half-maximum (FWHM) temporal width of T\textsubscript{pulse}$\approx50$~ps. The OPG is only powered by a single 5 V USB cable. The resulting optical pulse sequence is fed into a series of calibrated variable optical attenuators (VOAs) that adjust the power to generate a weak coherent state (WCS) that is coupled into the detector via a fiber polarization controller (PC) that is tuned to maximize detector counts. The maximum channel loss at which communication is demonstrated in this experiment, for $M=2^{21}$ and a mean number of photons incident on the detector per frame of $\lambda=0.02$, is $87.5$ dB.

The detection setup consists of an electrically-amplified cryogenic SNSPD connected to a time-tagging module (TTM). The SNSPD is biased using a constant electrical current that provides a good trade-off between the spontaneous thermal breaking of Cooper pairs within the SNSPD (i.e., dark counts) and the fraction of incident photons that are detected, namely the system detection efficiency SDE, which we symbolically denote as $\eta_{\mathrm{s}}$. A very small background count rate ($<1$~Hz) was observed in the experiment. We denote the combined dark and background count rate as DBCR. The SDE and DBCR are experimentally measured to be $\eta_{\mathrm{s}}=0.815 \pm 0.01$ and DBCR$\approx15$~Hz, respectively. A very-low-frequency (10 Hz) electrical clock is shared between the AWG and the TTM. The timestamps generated by the TTM are recorded via a computer interface and processed offline to analyze system performance.

In practical communication systems, such as in deep-space, background radiation power can vary considerably depending upon the sources present, the wavelength band of interest for communication, the aperture size and location of the receiver, and the optical path traversed by the transmitted signal \cite{Kevin2014,PsycheTechnical2023,Mohageg2022,Aboagye2021}. Several techniques exist for mitigation of background noise \cite{ESA_QPG,Lee2005,PsycheTechnical2023,Raymer:20,jarzyna2024photon}, a thorough review of which is beyond the scope of this paper. Here, we limit our attention to the case of low background noise, which is relevant for transmitters in the vicinity of small celestial bodies with weak albedo, as well as in the scenario of no celestial bodies in the line of sight between transmitter and receiver. A theoretical analysis of system behavior in the presence of large background noise is presented in Supplemental Document.

\section{Timestamps Analysis}
The detected PPM sequence is recovered from the recorded timestamps by sorting them into temporal bins that are generated by interpolating the $10$~Hz clock to the time slot rate. The relative delay between the optical signal and the clock arising from any path length difference is digitally compensated before performing the binning operation. For the purposes of error correction, a detected frame is considered a `valid' frame only if it contains exactly one count, i.e., if it lies within the transmitted PPM alphabet, and is compared with the corresponding transmitted frame and declared either an errored or a correctly-received frame. A detected frame is declared a `data frame erasure' if it contains either zero counts (referred to here as an empty frame erasure) or two or more counts at different locations in the frame (a multiple-count frame erasure), i.e., if it lies outside of the transmitted PPM alphabet. The latter primarily occurs when a dark/background count is registered in the same frame as a signal count, but additional lower-probability events include multiple dark/background counts or multiple counts from a finite-extinction-ratio signal pulse in one frame. Multiple counts in a single time slot is not possible here, since the SNSPD is not photon-number-resolving, as the time slot duration ($400$~ps) is much shorter than the SNSPD dead time ($60$~ns). A schematic illustration of the channel model is presented in Supplemental Document, where a detailed analysis of measured frame erasure and error ratios is also presented. The number of PPM frames recorded were $\approx$ 100k, 71k and 35k respectively for $M=2^{19},2^{20}$ and $2^{21}$, with 1000, 400 and 200 respectively being distinct. The latter numbers were only limited by the AWG memory.

The mean number of photons per frame incident on the detector, $\lambda$, is calculated from the measured empty frame erasure ratio by inverting the Poisson probability for an empty frame detection. Finally, the measured frame erasure and error ratios are used to compute achievable information rates using Reed-Solomon (RS) forward-error correction (FEC) codes in which $k$ information-carrying frames are encoded into an RS code of length $n=M-1$. We compute the PIE under the assumption of bounded distance decoding, i.e., we assume correct decoding only if $\xi>e+2t$, where $\xi$ is the minimum code distance, and $e$ and $t$ are the number of errors and erasures, respectively. We compute the largest possible $k=k^*$ such that the RS codeword decoding error probability has an upper bound of $10^{-6}$. The PIE is then given by PIE$(\lambda)=k^*\log_2(M)/(\lambda \, n)$. It was found that for the large code lengths chosen in this work, and for $\lambda\gtrsim0.1$, an RS codeword decoding error probability of $10^{-15}$ can be achieved at a PIE that is at most $2\%$ smaller.

\section{Results}
 \label{sec:results}
The average transmitter extinction ratio per time slot is defined as $\mathrm{ER_{slot}}=(M-1)R_s/(R_n-R_d)$ where $R_s$, $R_n$ and $R_d$ are the signal, noise, and SNSPD dark/background count rates respectively, where the noise rate includes all counts measured outside of the known transmitted time slot. The transmitter extinction ratio was measured to be $\mathrm{ER_{slot}}\approx90.8$~dB per slot. 

System behavior is assessed by comparing measured performance with a theoretical statistical model based on the Poisson number distribution of the WCS. Dark/background counts are assumed to have a mean Poisson rate equal to the experimentally-measured DBCR. Finally, the transmitted optical pulses are assumed to have an infinite extinction ratio. In such a model, all frame errors and multiple-count frame erasures are caused by dark/background counts. In other words, the theoretical system performance is dark/background-count-limited, referred to here as `DBCL'. As we will show, the measured system performance follows the theoretical DBCL model very closely, indicating that the high transmitter extinction ratio in the experiment ensures very minimal additional errors.

\begin{figure*}[htbp]
\centering
\includegraphics[width=\linewidth]{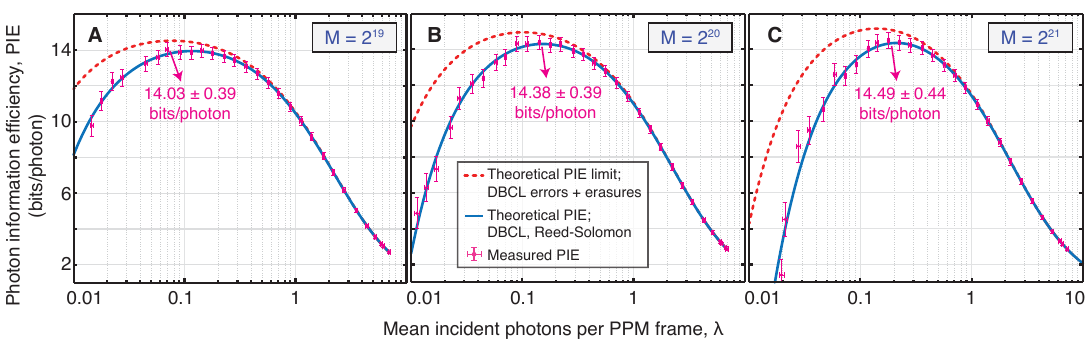}
\caption{Measured and theoretical dark-background-count-limited photon information efficiency (PIE) vs mean incident photon number per PPM frame, $\lambda$, for (a) $M=2^{19}$, (b) $M=2^{20}$, and (c) $M=2^{21}$. Dashed red curve: maximum achievable mutual information rate per incident photon for theoretical DBCL frame error and erasure probabilities. Solid blue curve: PIE achievable for such a channel upon employing RS coding. Both theoretical curves assume a transmitter source that produces infinite extinction ratio optical pulses. Magenta data points: measured PIE and its uncertainty estimates. All curves shown are computed for experimentally determined SDE and DBCR ($\eta_{\mathrm{s}}=0.815$, DBCR $=15$ Hz), allowing direct comparison of theoretical and measured PIEs.}
\label{fig:PIEvsPhotoncount}
\end{figure*}

Fig.~\ref{fig:PIEvsPhotoncount} indicates the experimentally-measured PIE (magenta data points with error bars) alongside the theoretical PIE limits. The dashed red curve depicts the maximum achievable mutual information rate per incident photon for a PPM channel with DBCL error and erasure probabilities (Equations~S6 and S7 in Supplemental Document), while the solid blue curve shows the PIE limit for such a channel upon employing RS coding. Both curves are computed for experimentally-determined SDE and DBCR values, namely $\eta_{\mathrm{s}}=0.815$ and DBCR~$=15$ Hz, allowing direct comparison of achievable and measured PIEs. The error bars for the measured PIE and $\lambda$ shown in Fig.~\ref{fig:PIEvsPhotoncount} indicate experimental uncertainty in the measured SDE and DBCR.

The experimentally-demonstrated PIE shows excellent overlap with the theoretically-computed PIE achievable with RS coding (solid blue curves), indicating that system performance is accurately described by an errors-plus-erasures channel model that employs RS coding, where the errors and erasures result solely from dark/background counts and channel attenuation. Interestingly, Fig.~\ref{fig:PIEvsPhotoncount} indicates that for large $\lambda$ ($\gtrsim0.6$), the performance approaches the errors-plus-erasures limit. This is explained as follows: for $\lambda\to0$, dark/background counts ``contaminate" only empty received frames, leading to frame errors. For $\lambda\gg1$, dark/background counts result only in multiple-count frame erasures rather than errors, resembling an erasure-only channel. In general, the bounded minimum distance decoding employed here is sub-optimal compared to maximum likelihood decoding. However, in the erasure-only channel, bounded distance decoding of RS codes provides optimal performance. As a result, the experimentally-demonstrated PIE approaches the errors-plus-erasures limit.

The maximum PIEs in Fig.~\ref{fig:PIEvsPhotoncount} are seen around $\lambda \sim 0.1$, corresponding to $14.03\pm0.39$, $14.38\pm0.39$, and $14.49\pm0.44$ bits per incident photon for PPM $2^{19}$, $2^{20}$, and $2^{21}$, respectively. When reported in terms of detected photons, they correspond to $17.21$, $17.64$, and $17.78$ bits per detected photon, and the associated incident energy per bit values are $9.13$, $8.91$, and $8.84$ zeptoJoules per bit. As the plots in Fig.~\ref{fig:PIEvsPhotoncount} show, there exists a broad range of average incident power on the detectors (i.e., a broad range of channel loss at fixed transmitter power) for which a PIE of over 10 bits/photon is achieved. To the best of our knowledge, these are the highest-ever demonstrated PIE numbers, or equivalently, the lowest energy ever detected at optical frequencies to receive one bit of information using a state of light compatible with high-loss channels.

\begin{figure*}[htbp]
\centering
\includegraphics[width=\linewidth]{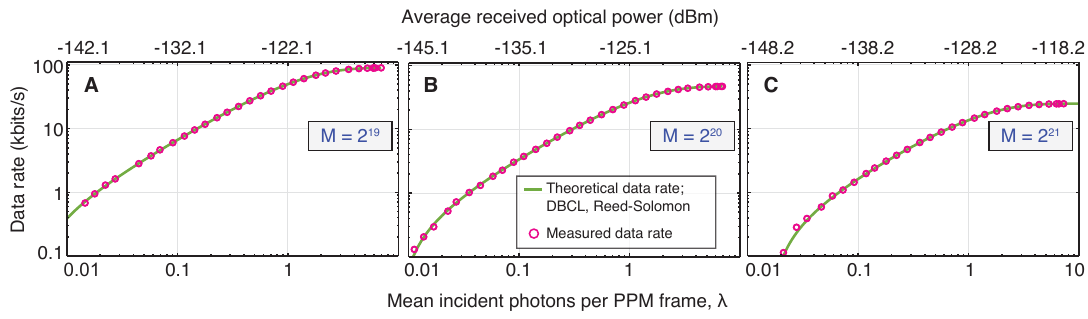}
\caption{Measured and theoretical data rate vs mean incident photon number per PPM frame, $\lambda$, and average incident optical power (top axes), for (a) $M=2^{19}$, (b) $M=2^{20}$, and (c) $M=2^{21}$. Solid green curve: theoretically computed data rate achievable for a PPM channel with DBCL error and erasure probabilities upon employing RS coding. Magenta circles: measured data rate. All curves shown are computed for experimentally determined SDE and DBCR ($\eta_{\mathrm{s}}=0.815$, DBCR $=15$ Hz), allowing direct comparison of theoretical and measured data rates.}
\label{fig:dataRatevsPhotoncount}
\end{figure*}

Fig.~\ref{fig:dataRatevsPhotoncount} depicts the measured and theoretical RS data rates, given by D$=k^*\log_2(M)/(n\mathrm{T_{frame}})$. Like the PIE, the measured data rate shows excellent agreement with the theoretically-computed data rate for DBCL frame erasure and error probabilities and RS coding. For $\lambda\lesssim1$, the data rate exhibits a close-to-linear dependence on the average incident optical power, whereas for $\lambda\gtrsim1$, the data rate saturates and asymptotically reaches the limit of the PPM frame rate. Such a behavior is explained by the strong reduction in frame erasure ratio in the $\lambda\gtrsim1$ regime, described in greater detail in Supplemental Document (Fig.~S6). These results illustrate the great flexibility in data rate offered by the large-$M$ platform over a very wide range of incident optical powers.

Finally, we note that high PIEs such as $10$~bits/photon at telecommunication (i.e., $\sim$1550~nm) wavelengths may be unattainable using single-photon avalanche diodes (SPADs) currently available commercially due to their significantly lower SDEs ($\lesssim0.25$) compared to SNSPDs ($\gtrsim0.8$) for similar DBCRs \cite{hamamatsuSPAD,excelitasSPCM}. However, some commercially available SPADs already offer SDEs exceeding $0.7$ at visible wavelengths \cite{excelitasSPCM}, with considerable progress being made at telecommunication wavelengths. This presents the exciting possibility that high-PIE communication in the infrared may be demonstrable using room-temperature detectors in the near future.

\section{Conclusions}
 \label{sec:conclusion}
In conclusion, we experimentally demonstrated a record photon information efficiency of $14.5$ bits per incident photon, or $8.84$~zeptojoules per bit, at $1550$~nm wavelength. This is achieved using a new low-complexity transmitter generating well-confined high-extinction-ratio short pulses on demand and a large PPM order. Such a high-PIE communication platform can enable bridging unprecedented path losses in power-constrained channels, in particular for deep-space exploration. Finally, we envision that the ability to extract information after severe attenuation can enable several other applications such as communication in turbulent media, increasing secrecy of transmission in fiber-based and free-space links, and improving the energy efficiency of conventional telecommunication networks.

\newpage
\onecolumngrid
\setcounter{equation}{0}
\setcounter{figure}{0}
\setcounter{table}{0}
\setcounter{section}{0}
\renewcommand{\theequation}{S\arabic{equation}}
\renewcommand{\thefigure}{S\arabic{figure}}

\newpage
\begin{center}
\textbf{\large Communicating at a record 14.5 bits per received photon through a photon-starved channel: Supplementary Information}
\end{center}

\section{Pulse-Position Modulation Scheme}

\begin{figure}[htbp]
\centering
\includegraphics[scale=1]{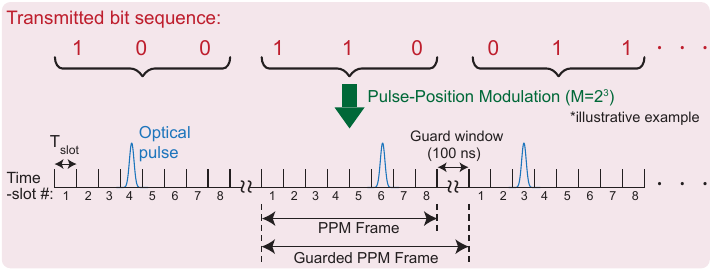}
\caption{Illustrative schematic depicting the pulse-position modulation (PPM) scheme employed in this work, for an example case of $M=2^3$. A computer-generated bit sequence is segmented into bit strings of length $m=\mathrm{log}_2(M)$, and each unique bit string is mapped to a unique temporal position of an optical pulse in one out of $M$ time slots. Consecutive PPM frames are separated by a guard window of 100 ns.}
\label{fig:PPMFrame}
\end{figure}

Fig.~\ref{fig:PPMFrame} depicts the pulse-position modulation (PPM) scheme employed in this experiment. A computer-generated random sequence of bits is converted into the PPM modulation format, in which exactly one time slot out of a set of $M=2^m$ time slots (referred to here as the PPM frame) contains an optical pulse while all other time slots are empty. The number of bits mapped to each PPM frame is given by $m=\mathrm{log}_2(M)$. As shown for the example case of $M=2^3$ in Fig.~\ref{fig:PPMFrame}, each of the $8$ different possible bit strings of length $m=3$ is mapped to a unique time slot number. A temporal $100$~ns-long `guard window' is added between two consecutive PPM frames, to minimize the probability of detector blocking that may arise when two consecutive PPM frames contain optical pulses that are closer than the dead time of the detector, which, for our detectors, is $\approx60$ ns. The dead time is defined as the minimum time interval required between successive photon arrivals for the detector to be able to detect both (up to the SDE). The combined time-domain symbol -- i.e. PPM symbol + guard window -- is referred to here as a `guarded PPM frame'. For the large values of $M$ described in this work, the temporal duration of the PPM frame is much larger (by at least four orders of magnitude) than the guard window. As a result, the addition of the guard window results in negligible penalty to the PIE and the data rate.

\begin{figure}[htbp]
\centering
\includegraphics[scale=1]{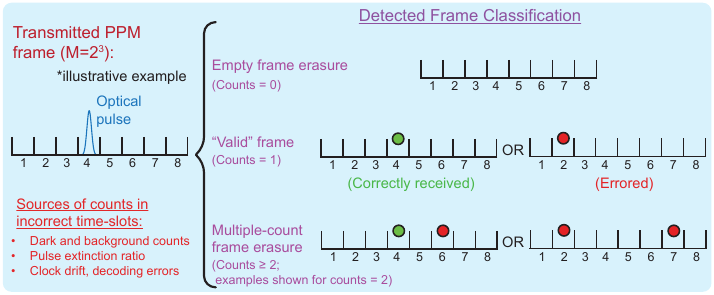}
\caption{Schematic illustrating possible detection events for an example case of $M=2^3$. Detected frames that contain zero counts are declared as empty frame erasures, while frames with two or more counts are declared multiple-count frame erasures. If a detected frame contains exactly one count, it is considered a `valid' frame, and is compared with the corresponding transmitted frame to assess frame errors.}
\label{fig:detEvents}
\end{figure}

We begin our discussion on detected frame classification by first noting that a `count', throughout this manuscript, refers to a detection event in which an electrical pulse is generated by the SNSPD driver and tagged by the TTM. When the temporal separation between consecutive signal photons incident on the detector is larger than the detector dead time, it can result in up to one count per signal photon incidence. In such a scenario, the measured number of `counts' is completely determined by the number of incident photons and the detection efficiency $\eta_s$. However, when the temporal separation between consecutive incident signal photons is less than the detector dead time, the incidence of multiple photons on the detector still only results in up to one count, since the detector is inactive during the dead time following a detection event. This phenomenon is known as ``detector blocking", which will be discussed in further detail later in this manuscript. Note that detector blocking can also be caused by dark/background counts. 

At the receiver, the detection of a PPM frame can result in the following three scenarios, as illustrated in Fig.~\ref{fig:detEvents}: 1) an empty frame detection, 2) a detected frame with exactly one count, and 3) a detected frame with two or more counts. For the purposes of error-correction coding, the first and the third possibilities are declared to be data frame erasures since they fall outside of the alphabet of transmitted symbols. An `empty frame erasure' results when neither a dark or background count nor a photon from the transmitter is detected in the frame, whereas a `multiple-count frame erasure' occurs when two or more time slots contain counts, coming either from signal photons or dark/background noise. Two such examples are represented in Fig.~\ref{fig:detEvents}. Finally, if a detected frame contains exactly one count, it is considered a valid frame, and is compared with the corresponding transmitted frame to assess frame error performance of the system.

\begin{figure}[htbp]
\centering
\includegraphics{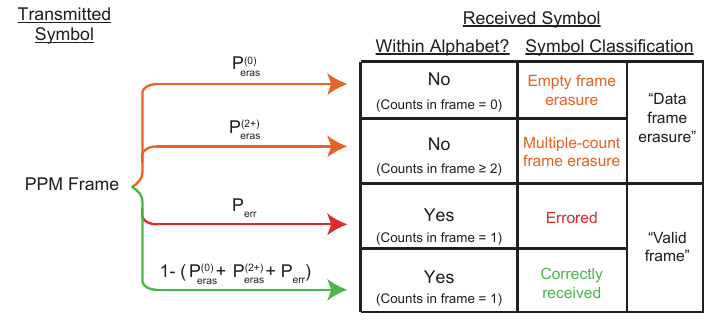}
\caption{Schematic depicting PPM erasure channel diagram. Each detected PPM frame is categorized into in one of four outcomes based on the number of counts measured in the frame.}
\label{fig:channelDiag}
\end{figure}

Fig.~\ref{fig:channelDiag} illustrates the error and erasure channel diagram corresponding to the detection events described in Fig.~\ref{fig:detEvents}. When the transmitted symbol is a PPM frame, there are four possibilities for the received symbol. As described above, the received PPM frame can be either empty or contain counts in multiple time slots, with probabilities denoted by $\mathrm{P^{(0)}_{eras}}$ and $\mathrm{P^{(2+)}_{eras}}$, respectively. Since both are outside of the transmitted PPM alphabet, they are classified for error-correction purposes as ``data frame erasures". The remaining two possibilities include the received PPM frame containing a count in exactly one time slot. Since the received symbol is part of the transmitted PPM alphabet, such frames are classified as ``valid" frames. Valid frames are either errored or correctly received frames, with probabilities denoted by $\mathrm{P_{err}}$ and $1-(\mathrm{P^{(0)}_{eras}}+\mathrm{P^{(2+)}_{eras}}+\mathrm{P_{err})}$, respectively. The analytical expressions for the dark-background-count-limited frame erasure and error probabilities are described later, in Equations \ref{eq:PEras} and \ref{eq:PErr}.


\section{Measurement of Detection Efficiency and Dark \& Background Count Rate}

The SNSPD used in the detection setup is biased using a constant electrical current, denoted as $\mathrm{I_{bias}}$, such that it operates in the superconducting regime just below the threshold for transition to a resistive state. At such an operating point, the energy absorbed from a single photon is sufficient to break superconductivity for a short period of time (referred to as the detector dead time) and produce a rapid voltage change, resulting in a detection event~\cite{you2020superconducting}. The current $\mathrm{I_{bias}}$ is adjusted to provide a good trade-off between the generation of spontaneous thermal breaking of Cooper pairs within the SNSPD (dark counts) and the fraction of incident photons that are detected, namely the system detection efficiency SDE, which we symbolically denote as $\eta_{\mathrm{s}}$.

We measure the SDE and DBCR of the SNSPD by following the method presented in Refs.~\cite{Marsili2013,Vyhnalek2018}. A continuous wave (CW) DFB laser operating at 1549.9 nm is attenuated using a bench-top variable optical attenuator (VOA) to $\approx$ -50 dBm. The output power is measured using a calibrated, high accuracy and high resolution bench-top optical power meter. A separately calibrated (second) bench-top VOA is used to further attenuate the CW beam to $\approx$ -111 dBm so as to produce a count rate around $\approx$ 60,000 counts per second on the detector. We then measure the number of detected photon counts over a long averaging time (60 s) for a range of SNSPD bias currents. The SDE is extracted by normalizing the mean number of measured counts per second to the input optical power measured beforehand (converted to a mean photon number at the laser wavelength).

In order for the SDE to be accurately measured as a simple ratio of the measured counts and average incident power, it is important to minimize the probability of detector blocking, i.e., one photon ``blocking" the detection of a second photon that arrived within the dead time of the detector after the first one. This is done by choosing the order of magnitude for the measured count rate (in this case, $\approx$ 60,000 counts per second) such that the average inter-photon arrival time (in this case, $\approx$ 20 $\mu$s) is much larger than the dead time of the SNSPD, which was measured separately to be $\approx60$~ns, and which we define here as the minimum temporal spacing between consecutive photon arrivals for the detector to be able to detect both (up to the SDE).

\begin{figure}[htbp]
\centering
\includegraphics{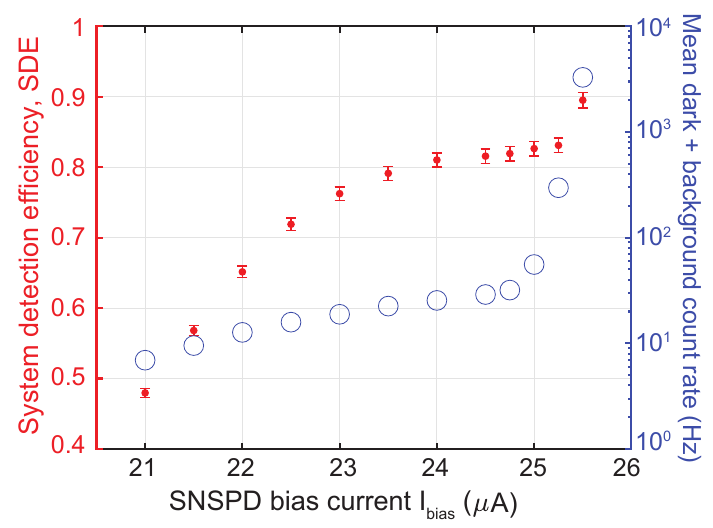}
\caption{Characterization of SNSPD: SDE and DBCR vs detector bias current. The error bars depicted correspond to one Poissonian distribution standard deviation in the measured count rates and an additional 0.1 dB error in the measured calibration optical power.}
\label{fig:DetEff}
\end{figure}

Separately, we also measure the combined dark + background count rate, DBCR, as a function of SNSPD bias current $\mathrm{I_{bias}}$. This is done by turning off the laser transmitter and averaging the counts measured over $\approx60$~s. The dark counts were measured by covering the input port of the SNSPD with a metallic cap, and the background counts were measured by connecting the fiber to the SNSPD input port, turning off the transmitter laser and carefully isolating the experimental setup from ambient light in the laboratory. No optical bandpass filters were used. The results of this characterization are shown in Fig.~\ref{fig:DetEff}. The error bars depicted in the SDE plot are assessed by considering Poissonian standard deviation error in the measured count rates as well as a possible 0.1 dB error in the measurement of the calibration input optical power arising from disconnecting/reconnecting fiber optical connectors during the measurement.

This experiment was carefully performed multiple times in order to ensure repeatability of the measured SDE and DBCR. The trends depicted in Fig.~\ref{fig:DetEff} agree with the SDE and DBCR vs bias current trends previously reported in the literature, such as in \cite{Vyhnalek2018}. A bias current of $I_b \approx$ 24.5 $\mu$A was chosen for the PPM detection experiments as it provided a high SDE $\approx 0.815 \pm 0.01$ while maintaining a low mean DBCR of $\approx 15$ Hz.

\section{Measurement of $\lambda$}

Having properly characterized the SNSPD as well as the transmitted optical pulse, we now move to measuring the mean number of photons incident on the detectors per PPM frame, $\lambda$, which is critical to the accurate assessment of the photon information efficiency. To this end, we adopt a methodology similar to the one discussed in \cite{Farr2013}, namely inverting the experimentally\-measured \textit{empty} frame erasure probability. Modeling the system performance using Poisson statistics, the probability of detecting an empty frame is given by:


\begin{equation}
    \mathrm{P^{(0)}_{eras}}(\lambda)=e^{-\eta_s\lambda-\lambda_d}
    \label{eq:Pempty}
\end{equation}

Experimentally, we measure the fraction of detected frames that are empty, denoted here as $\mathrm{F^{(0)}_{eras}}$. In the presence of sufficient statistics in the experimental data, we set the left-hand-side of Equation~\ref{eq:Pempty} equal to $\mathrm{F^{(0)}_{eras}}$, and invert the expression to recover $\lambda$:


\begin{equation}
    \lambda = \frac{-(\mathrm{ln(F^{(0)}_{eras})}+\lambda_d)}{\eta_{\mathrm{s}}}
    \label{eq:lambda}
\end{equation}
where $\lambda_d$ is the DBCR normalized to the duration of one guarded PPM frame, and $\eta_s$ is the SDE of the SNSPD.

We note that we also performed separate measurements, not described here, in which we estimated $\lambda$ by performing a careful measurement of the input average optical power using a highly sensitive optical power meter. Although the results of those measurements broadly agree with the $\lambda$ measured using the probability inversion method, the latter was observed to be more repeatable.

\section{Pulse Shape and Extinction Ratio Measurements}

The transmitter employed in this work consists of a single directly-modulated laser system, unlike the cascade of optical devices reported previously. Specifically, the OPG used here gain-switches a DFB laser to produce short optical pulses with a high extinction ratio, which is necessary to achieve DBCR-limited performance. Here, we follow the definition of ER presented in \cite{Farr2013}:

\begin{equation}
    \mathrm{ER} = \frac{R_s}{R_n - R_d}
    \label{eq:ER}
\end{equation}

where $R_s$, $R_n$ and $R_d$ are the measured signal, noise and dark count rates respectively. ``Noise rate" $R_n$ is defined as the fraction of detected photon counts that lie in all time slots other than the transmitted pulse time slot. This includes dark/background counts as well as spurious counts arising from any residual emissions from the transmitter outside of the transmitted time slot. We define the average extinction ratio per time slot as follows:

\begin{equation}
    \mathrm{ER_{slot}} = \frac{(M-1)R_s}{R_n-R_d}
   \label{eq:ERSlot}
\end{equation}

where $M$ is the total number of time slots in a PPM frame. The extinction ratio per slot provides a measure of how large of a PPM sequence a given pulsed optical transmitter can support. With a transmitter with a large ER\textsubscript{slot}, one can support a large PPM order $M$ without the empty time slots being corrupted, whereas a smaller ER\textsubscript{slot} implies the presence of a larger number of photons outside of the desired time slots, which will ultimately hinder achieving high-photon-efficiency communication.

\begin{figure}[htbp]
\centering
\includegraphics{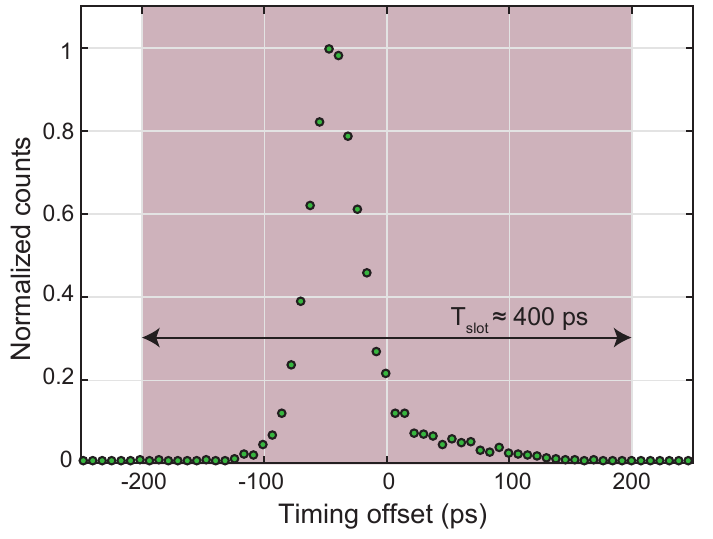}
\caption{Shape of the transmitted optical pulse reconstructed by overlaying arrival time\-stamps of photons detected in known time slots over a long acquisition time. The shaded portion represents T\textsubscript{slot} $\approx$ 400 ps, the duration of one PPM time slot. The optical pulse, generated by directly modulating a DFB laser, consists of weak tails that are confined well within this time slot, thereby providing sufficient extinction ratio to achieve near dark-background-count-limited detection.}
\label{fig:PulseShape}
\end{figure}

In order to experimentally characterize the ER of the optical pulse source, we first transmit a single PPM frame repeatedly, and reconstruct the shape of the optical pulse from the detected photon arrival time-stamps measured over a long acquisition period. By overlaying successive detected frames, we reconstruct the temporal shape of the optical pulse, as shown in Fig.~\ref{fig:PulseShape}. The PPM order chosen for this measurement ensures that the temporal spacing between consecutive transmitted pulses was much larger than the detector dead time of 60 ns, thereby minimizing the probability of detector blocking.

As one can ascertain visually from Fig.~\ref{fig:PulseShape}, although the gain-switched optical pulse has a weak but observable tail, it is well confined within the T\textsubscript{slot} $\approx$ 400 ps PPM time slot. By measuring the total counts received inside and outside the transmitted time slots, the ER\textsubscript{slot} is estimated. Specifically, by adjusting the mean incident photon number per frame $\lambda$ to be $\approx0.1$ over a long measurement period, we measure 31,746 counts in the transmitted time slot, and a total of 148 noise counts, of which 28 are assessed, based on arrival timestamps, to be arising from weak pulse tails outside of the transmitted time slot, which is expected to be the most significant contribution to noise counts from the transmitter. The resulting net extinction ratio, i.e., the ratio of power in the transmitted time slot to the power emitted by the transmitter in the remainder of the frame, is ER $\approx$ 30.5 dB. When normalized to one time slot, for $M=2^{20}$, this results in an average extinction ratio of ER\textsubscript{slot} $\approx$ 90.8 dB. 

We note that the ER\textsubscript{slot} achieved with the transmitter presented here is significantly better compared to the scheme described in \cite{Farr2013}. This can be attributed to the fact that the directly modulated laser here is reverse biased throughout the empty time slots, whereas the scheme presented in \cite{Farr2013} relies upon external modulation of a CW laser with a gain-switched semiconductor optical amplifier (SOA) cascaded with a time-synchronized electro-optic modulator. The transmitter system described here not only provides a superior ER\textsubscript{slot}, but it also is a more compact, simpler and more energy-efficient system in comparison to cascaded external optical modulators, which makes such a source very practical, and ideal for deep-space missions that operate on a minimal supply of electrical power.

\section{Frame Error and Erasure Rates}

In this section, we assess the accuracy of the theoretical statistical model in which system behavior is described by Poisson probabilities of the WCS and dark/background counts, by comparing the measured frame erasure and error rates to the theoretical DBCL probabilities.

As described in the Main Text and in Figs.~\ref{fig:detEvents} and \ref{fig:channelDiag}, a detected frame is declared a data frame erasure if it is either empty (empty frame erasure) or contains two or more counts (multiple-count frame erasure). The latter occurs for either a combination of a dark/background and signal counts in the same frame or for multiple dark/background counts in the same frame. The theoretical DBCL data frame erasure probability is therefore given by:


\begin{equation}
    \mathrm{P_{eras}}(\lambda)=\mathrm{P^{(0)}_{eras}}(\lambda)+\mathrm{P^{(2+)}_{eras}}(\lambda)
    \label{eq:PErasBasic}
\end{equation}

Since all received frames with either zero or two or more counts are declared erasures, Equation~\ref{eq:PErasBasic} can be conveniently re-expressed as one minus the probability of receiving a frame with exactly one count:

\begin{multline}
    \mathrm{P_{eras}}(\lambda)=1-\biggl(\underbrace{\lambda_d e^{-(\lambda_d+\eta_s\lambda)}}_{\substack{\text{1 dark/backg. count},\\\text{0 signal count}}}+\underbrace{(1-e^{-\eta_s\lambda})e^{-\lambda_d}}_{\substack{\text{0 dark/backg. count},\\\text{1 signal count}}}\biggr)\underbrace{\biggl(\frac{M-\delta}{M}\biggr)}_{\substack{\text{counts}\\\text{separation } > \delta}}-\\
    \biggl(\underbrace{\lambda_d e^{-\lambda_d}}_{\substack{\text{1 dark/backg. count,}\\ \text{arb. signal counts}}} +\underbrace{(1-e^{-\eta_s\lambda})}_{\substack{\text{1 signal count,}\\\text{arb. dark/backg.}\\\text{counts}}}\biggr)\underbrace{\frac{\delta}{M}}_{\substack{\text{counts}\\\text{separation }<\delta}}
   \label{eq:PEras}
\end{multline}

where $\lambda_d$ is the DBCR normalized to the duration of one guarded PPM frame, $\delta$ is the number of time slots corresponding to the dead time of the SNSPD, and the sum of all terms inside the square brackets is the probability of receiving a frame with exactly one count. The first two highlighted terms inside the square brackets on the right-hand-side of Equation~\ref{eq:PEras} denote the Poisson probability of receiving either one signal count and zero dark/background counts per frame, or vice versa, respectively. The multiplicative factor $\frac{M-\delta}{M}$ denotes the fact that the two probabilities are computed for the case of counts separated by more than the detector dead time. The second group of highlighted terms account for the possibility of exactly one (either signal or dark/background) count in a frame being followed by any arbitrary number of either signal or dark/background \textit{incidences} separated by less than the detector dead time, in which case the latter incidences do not result in counts due to detector blocking.

Finally, we note that the most general expression for the probability of producing one \textit{signal count} in a frame is given by $1-e^{-\eta_s\lambda}$, i.e., one minus the Poisson probability of zero signal counts, and not by the Poisson probability of exactly one detected signal count $\eta_s\lambda e^{-\eta_s\lambda}$. This is because we model the transmitter to have an infinite extinction ratio, i.e., all signal photons incident on the detector are modeled to be confined to the same time slot. As a result, any value of $\lambda$ can only produce either zero or one count. This is in contrast with the dark/background counts, which are modeled to be temporally spread throughout the entire frame.


Similarly, the theoretical DBCL frame error probability is given by:

\begin{equation}
    \mathrm{P_{err}}(\lambda)=\left[\lambda_d e^{-\eta_{\mathrm{s}}\lambda-\lambda_d}\biggl(\frac{M-\delta}{M}\biggr)+\lambda_d e^{-\lambda_d}\frac{\delta}{M}\right]\biggl(\frac{M-1}{M}\biggr)
    \label{eq:PErr}
\end{equation}

where the first term denotes the probability of the absence of signal photons and the presence of exactly one dark count, and the second term indicates the fact that if a dark count were to occur approximately $T_{DT}$ (i.e. the detector dead time) before a signal photon would arrive at the detector in the wrong time slot, then it would block the detection of the signal photon and result in a frame error. The multiplicative term $\frac{M-1}{M}$ accounts for the fact that a single-count frame results in an error only when the count is in the wrong time slot. Once again, the impact of $\delta$ in Equation \ref{eq:PErr} is very minimal ($\ll0.01\%$) since $\delta\ll M$ in this work.

\begin{figure}[htbp]
\centering
\includegraphics[width=\linewidth]{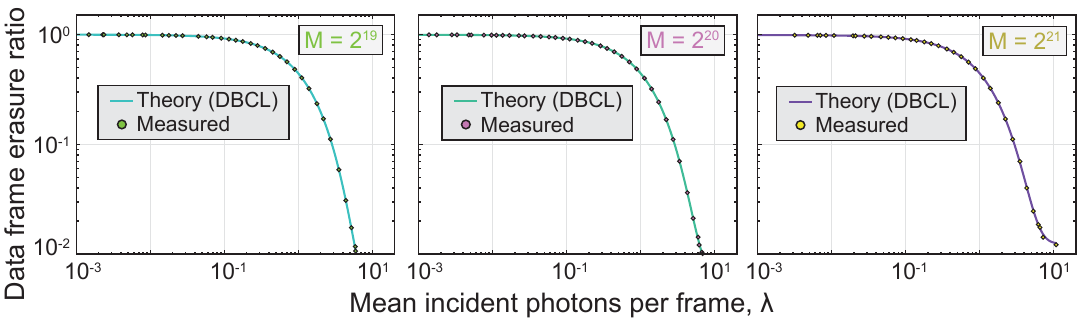}
\caption{Comparison of theoretical dark-background-count-limited (DBCL) and experimentally measured PPM frame erasure ratio, for PPM $2^{19}$, $2^{20}$ and $2^{21}$ with T\textsubscript{slot} $\approx$ 400 ps. The theoretical curves are plotted for a detection efficiency $\eta_{\mathrm{s}} = 0.815$ and a dark/background count rate (DBCR) of 15 Hz.}
\label{fig:FrameEras}
\end{figure}

\begin{figure}[htbp]
\centering
\includegraphics[width=\linewidth]{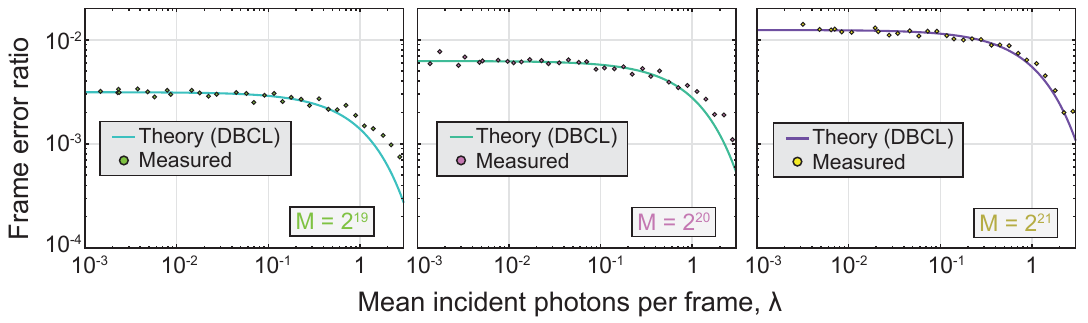}
\caption{Comparison of theoretical dark-background-count-limited (DBCL) and experimentally measured PPM frame error ratio, for PPM $2^{19}$, $2^{20}$ and $2^{21}$ with T\textsubscript{slot} $\approx$ 400 ps. The theoretical curves are plotted for a detection efficiency $\eta_{\mathrm{s}} = 0.815$ and a dark count rate (DBCR) of 15 Hz. The slight deviation of experimental observation from DBCL performance for $\lambda\gtrsim$1 can be attributed to the lack of sufficient frame error statistics at such high $\lambda$, with possible minor contributions from the finite extinction of the pulse tails outside of the transmitted time slot.}
\label{fig:FrameErr}
\end{figure}

We now compare the probabilities described in Equations~\ref{eq:PEras} and \ref{eq:PErr} to experimentally measured data frame erasure and frame error ratios. Fig.~\ref{fig:FrameEras} shows the comparison of data frame erasure ratios for PPM $2^{19}$, $2^{20}$ and $2^{21}$. Excellent agreement is observed between theory and experiment. Fig.~\ref{fig:FrameErr} shows a comparison of measured frame error ratios -- i.e., the fraction of \textit{all} detected frames that are declared as errors. We once again note excellent agreement, indicating that system performance is accurately captured by the theoretical DBCL model in Equations~\ref{eq:PEras} and \ref{eq:PErr}. The close agreement of experimentally measured performance with a theoretical model in which the optical pulses are assumed to have an infinite extinction ratio indicates that the extinction ratio of the transmitter used is sufficiently high for the values of $M$ chosen in this work.

The slight excess frame error ratio observed for $\lambda\gtrsim1$ in Fig.~\ref{fig:FrameErr} can be attributed to the lack of sufficient frame error statistics at such high $\lambda$. Minor contribution from the finite extinction of the pulse tails outside of the transmitted time slot may also be expected. Nonetheless, we note that these values of $\lambda$ are far separated from the regime of high-PIE operation that is the focus of the current work.

\section{Experimental Uncertainty Analysis}
\subsection*{Experimental Uncertainty in $\lambda$}

The experimental uncertainty level on $\lambda(\eta_{\mathrm{s}},\lambda_d)$ can be estimated by performing error propagation analysis:

\begin{equation}
    \biggl(\frac{\sigma_\lambda}{\lambda}\biggr)^2=\biggl(\frac{\partial\lambda}{\partial\eta_{\mathrm{s}}}\biggr)^2\sigma^2_{\eta_{\mathrm{s}}}+\biggl(\frac{\partial\lambda}{\partial\lambda_d}\biggr)^2\sigma^2_{\lambda_d}
\end{equation}

which, using Equation \ref{eq:lambda}, is rewritten as:

\begin{equation}
    \biggl(\frac{\sigma_\lambda}{\lambda}\biggr)^2=\biggl(\frac{\sigma_{\eta_{\mathrm{s}}}}{\eta_{\mathrm{s}}}\biggr)^2+\biggl(\frac{\sigma_{\lambda_d}}{\mathrm{ln(F_{empty})}+\lambda_d}\biggr)^2
\end{equation}

The temporal distribution of dark and background counts can be described by Poissonian statistics with a mean rate of $\lambda_d$ per frame. With ``infinite" statistics (i.e. an ``infinite" observation time), the measured mean (i.e. sample mean) $\lambda_d$ converges to the true average number of dark counts per frame, say $\lambda_d^\infty$, determined by the SNSPD's bias current as well as background counts, and $\sigma^2_{\lambda_d}=0$. However, due to the finite duration of data acquisition, the sample mean $\lambda_d$ differs from the true statistical mean $\lambda_d^\infty$, and the variance of sample mean is nonzero: $\sigma^2_{\lambda_d}\neq0$.

For a sufficiently large sample size, the sample mean variance can be estimated by dividing the sample variance of the random variable by the number of samples. If $N_{f}$ denotes the total number of PPM frames observed during an experiment, and if $q$ is a discrete random variable that describes the number of dark counts observed in each PPM frame (i.e. $q$ can take the values $0,1,2...$ for any given PPM frame), then the sample mean variance is given by:

\begin{equation}
    \sigma^2_{\lambda_d}=\frac{\sigma^2_q}{N_f}
    \label{eq:SEV_lam-d}
\end{equation}

We estimate $\sigma^2_{\lambda_d}$ by using this relation for the same dataset of dark/background counts recorded by the detector with no signal input that was used in Fig.~\ref{fig:DetEff}. $N_f$ is estimated as the total dark/background count acquisition time (i.e. $\approx60$~s) divided by the temporal duration of each frame $\mathrm{T_{frame}}$, $N_f\approx286,000$.


\subsection*{Experimental Uncertainty in PIE}

The experimental uncertainty margins along the $y$ axis of Main Text Fig.~4 are estimated by performing an error propagation analysis for the PIE. The PIE is given by the expression $\mathrm{PIE}=k^*\mathrm{log}_2(M)/(\lambda n)$, where, as described in the Main Text, $k^*/n$ denotes the highest RS code rate for which the RS codeword error probability has an upper bound of $10^{-6}$. The sources of uncertainty, in this case, are $\lambda$ and $k^*$. $\sigma^2_\lambda$ is estimated as described above. The experimental uncertainty in $k^*$ results from the finiteness of dark count statistics -- and by extension finiteness of frame error statistics -- gathered during the experiment. In other words, because the measured sample mean number of dark counts per frame $\lambda_d$ has a non-zero variance as discussed above (see Equation \ref{eq:SEV_lam-d}), so does the resulting frame error ratio. This, in turn, causes an uncertainty in the optimized RS code rate $k^*$. 

The larger vertical (i.e. PIE) error bars for lower $\lambda$ values seen in Main Text Fig.~4 can be explained as follows. When the measured frame error ratio is primarily determined by dark counts (i.e. at very low $\lambda$), the finite statistics of the sample mean variance $\sigma^2_{\lambda_d}$ has a much greater contribution. For larger $\lambda$, on the other hand, an increasing fraction of frame errors are caused by the finite extinction of the optical pulse (i.e. the pulse tails). Additionally, for larger $\lambda$, the RS code rate $k^*/n$ is higher, owing to lower empty frame erasures, leading to lower fractional uncertainty.

\section{Hybrid-PPM Transmitter for Deep-Space}
\label{Suppl:hybridPPM}
In order to further illustrate the suitability of a PPM-based transmitter for downlink deep-space optical communication, we theoretically analyze a ``hybrid-PPM" scheme in which the transmitter employs a varying PPM order that maximizes data rate at any given distance of the spacecraft from Earth. We assume a transmitter configuration that consists of a single-frequency laser directly modulated by an OPG, akin to the transmitter described in the Main Text, followed by a low-noise high-gain optical amplifier operating in the saturation regime that provides a fixed output average optical power. For simplicity, we model the transmitted optical pulses to have an infinite extinction ratio, which then lets us use Equations \ref{eq:PEras} and \ref{eq:PErr} to conveniently compute system performance.

We assume that the spacecraft, including its optical communication module, is powered by solar panels. We model the impact of the $1/r_s^2$ reduction, where $r_s$ is the distance of the spacecraft from the Sun, in electrical power generated by the solar panels by proportionally dialing down the optical amplifier gain and maintaining a fixed electro-optic efficiency. The transmitted optical signal from the spacecraft to Earth exhibits its own inverse-squared attenuation, as it varies as $1/r_e^2$ where $r_e$ is the distance from an Earth-based receiver. The total photon flux reaching the receiver is computed by assuming a fixed receiver ``antenna" size of 2.28 m. Finally, we assume RS coding, same as described in the Main Text.

In order to maximize the data rate at any given distance from an Earth-based receiver, such a transmitter would employ a lower PPM order when the spacecraft is close to the Earth and the Sun, where sufficient electrical power is generated by the solar panels and the mean number of photons per frame incident on the receiver, $\lambda$, is sufficiently high. As the spacecraft ventures farther away from the Earth and the Sun, the transmitter switches to progressively higher PPM orders that are more photon efficient. Using this scheme, a single transmitter can achieve high data throughput close to Earth, and more importantly, maintain reasonable data rates even at very long distances from Earth thanks to the very high PIE offered by large PPM orders. The flexibility that is provided by such a transmitter in achieving a broad range of data rates and PIE was experimentally demonstrated by us recently in Ref.~\cite{Guo2024CLEO}.

Table \ref{table:hybridPPMParams} summarizes the parameters used in this numerical study. The PPM slot width, guard window between consecutive PPM frames, receiver DBCR, and receiver SDE are all taken to be the same parameters employed in the Main Text. Several other parameters were chosen to match system parameters in Psyche's optical communication demonstration \cite{PsycheTechnical2023} and other systems-level lab demonstrations \cite{Dailey2019HighPowerTransm}. The transmitter is assumed to operate at an output optical average power of 4~W when sufficient power is available from the solar panels. The amplifier gain, and therefore the output average optical power, is proportionally reduced as the power generated by the solar panels reduces at farther distances. A fixed electro-optic efficiency -- i.e. the efficiency of converting electrical to optical power -- of 0.2 is assumed. The solar panel electrical power assumed here is similar to the power generated by Psyche's solar panels~\cite{nasa2023psyche}.

\begin{table}[h!]
\centering
\begin{tabular}{|c|c|} 
\hline
Parameter & Value \\ [0.5ex] 
\hline\hline
Power generated by solar panels at 3 AU from the Sun & 2000 W\\ 
\hline
Solar panel power dependence on $r_s$ & 1/$r_s^2$\\
\hline
Transmitter wavelength & 1550~nm\\
\hline
Transmitter electro-optic efficiency& 0.2\\
\hline
Transmitter antenna diameter & 0.22 m\\
\hline
Max. transmitter average optical power & 4~W\\
\hline
DBCR at receiver & 15~Hz\\
\hline
Receiver SDE & 0.815\\
\hline
PPM slot width, $\mathrm{T_{slot}}$ & 400~ps\\
\hline
Guard window between consecutive PPM frames & 100~ns\\
\hline 
Receiver (photon collection) antenna diameter & 2.28~m\\ [1ex]
\hline 
\end{tabular}
\caption{Parameters for numerical simulation of hybrid-PPM scheme}
\label{table:hybridPPMParams}
\end{table}

\begin{figure}[htbp]
\centering
\includegraphics{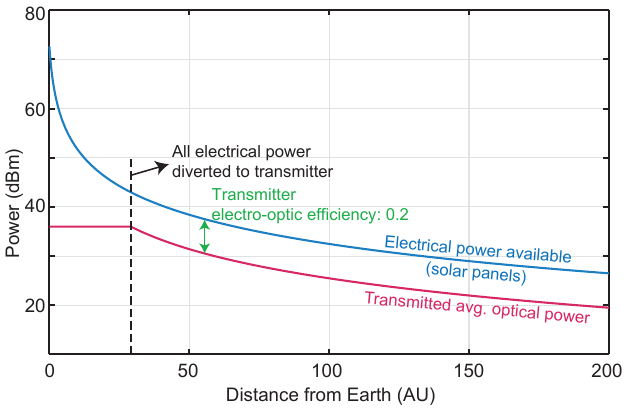}
\caption{Variation of electrical power generated by solar panels, and the resulting variation of output average optical power, with distance from earth. AU: Astronomical Units}
\label{fig:solarPanelPower}
\end{figure}

Fig.~\ref{fig:solarPanelPower} depicts the variation of electrical power generated by the solar panels. Specifically, it exhibits a $1/r_s^2$ dependence, where $r_s$ is the distance from the Sun. For the sake of uniformity, all quantities are plotted vs distance from Earth, where we make the simplifying assumption that the Sun, the Earth and the spacecraft all lie in a straight line. The magenta line in Fig.~\ref{fig:solarPanelPower} depicts the resulting variation in average optical power. The transmitter emits at a constant 4~W output power up to $\approx29$~AU from Earth, where the total electrical power generated by the solar panels equals the electrical power consumed by the 4~W transmitter. For distances beyond $29$~AU, we assume that the spacecraft temporarily shuts off power to all other instruments at the time of communicating with Earth so as to maximize data rate.

\begin{figure}[htbp]
\centering
\includegraphics{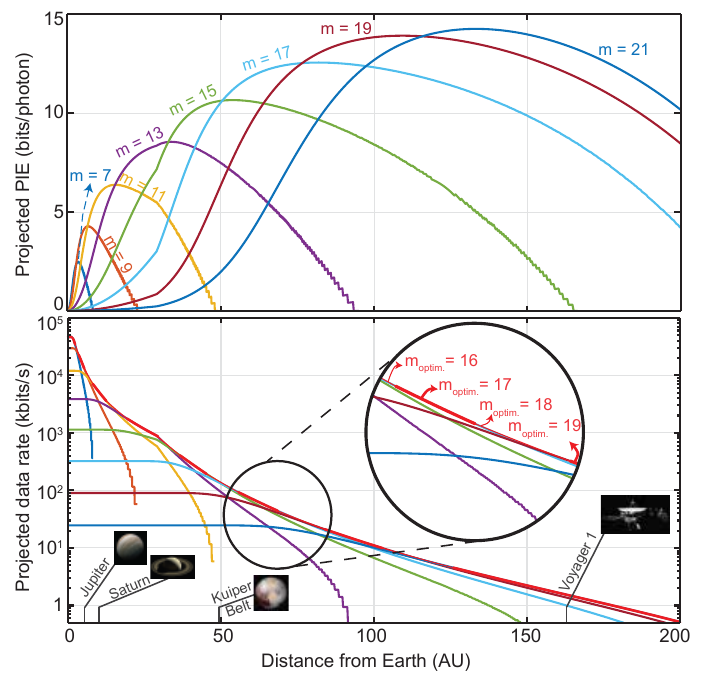}
\caption{Theoretical performance of a hybrid-PPM transmitter. (Top) Photon information efficiency of various PPM orders vs distance from Earth. (Bottom) Data rate offered by various PPM orders vs distance from Earth. The transmitter switches to progressively lower ``optimal" PPM orders, denoted by $\mathrm{m_{optim.}}$, maximizing data rate at all distances. The change in $\mathrm{m_{optim.}}$ is graphically depicted by the alternating thicknesses in the red curve.}
\label{fig:hybridPPMPerf}
\end{figure}

Fig.~\ref{fig:hybridPPMPerf} (Top) shows the PIE of various logarithmic PPM orders (denoted by $\mathrm{m=log_2(M)}$) as a function of distance from Earth. The average optical power emitted by the transmitter is assumed to vary is depicted in Fig.~\ref{fig:solarPanelPower}. As expected, smaller $m$ produces lower PIE closer to Earth, while larger $m$ produces higher PIE at larger distances. Fig.~\ref{fig:hybridPPMPerf} (Bottom) shows the corresponding data rate offered by the various PPM orders. Closer to Earth, smaller $m$ is favorable as it can enable high data rates approaching 100 Mbits/s for the parameters used here. By switching to progressively larger $m$ as the distance from Earth increases, data rate can be maximized at all distances by leveraging the higher PIE offered by larger $m$. Furthermore, larger $m$ enables a longer range of communication, since for the same average optical power at the transmitter, the energy per pulse is higher thanks to the lower repetition rate (i.e., frame rate).

The staircase-like features exhibited in the PIE and data rate plots arise from the fact that error-correcting codes, including the RS codes employed here, always have an \textit{integer-valued} dimension $k$. For a given code length (which, here, is related to the PPM order $M=2^m$ as $n=M-1$), the optimal dimension $k$ that maximizes data rate is constant over a range of attenuation (i.e. distance).

The projected data rate surpasses the $\sim10$~Mbits/s goal for communication from Jupiter/Saturn laid out in Ref. \cite{Edwards2023_NASA}. Furthermore, for comparison with existing communication systems at much larger distances, NASA's New Horizons spacecraft returned the now-iconic Pluto photograph (shown as inset in Fig.~\ref{fig:hybridPPMPerf}) at a data rate of approximately 2~kbits/s \cite{nasaNewHorizons}. At the same distance, the hybrid-PPM transmitter shows a projected data rate of $\approx138$ kbits/s, two orders of magnitude higher. Similarly, Voyager 1 currently operates at a data rate of 160 bits/s at the current distance of $\approx163$ AU. At the same distance, the hybrid-PPM transmitter can offer a data rate of $\approx1.5$~kbits/s, an order of magnitude higher.

It must be emphasized that the results presented in this Section are theoretical projections that do not take into account real-world conditions such as high DBCR arising from sky radiance and celestial noise sources, imperfect clock recovery, and pointing and coupling losses at the receiver. Additional losses are only expected to shift the curves in Fig.~\ref{fig:hybridPPMPerf} to the left along the distance axis, while noise in the received data degrade system performance at any given spacecraft to Earth distance, as discussed in the following Section. Nevertheless, Fig.~\ref{fig:hybridPPMPerf} illustrates the flexibility afforded by the PPM modulation format over a vast range of distances, and the benefit of employing a large $M$ to extend the span of the communication link.

\section{System Performance at High DBCR}
\label{Suppl:HighDBCR}

So far, we have focused our attention to an idealized case in which dark/background noise power is very low. In a real-life communication link, either in deep-space or in other terrestrial applications, however, background radiation is expected to be higher than in the lab experiment discussed in this work. In the deep-space context, background noise from stray photons, emitted by other sources in the communication path such as planetary albedo, solar noise, sky radiance, etc., entering the receiver can result in a large DBCR~\cite{Kevin2014,PsycheTechnical2023,Mohageg2022,Aboagye2021}. A thorough review of background noise mitigation techniques is beyond the scope of this paper, but the theoretical formalism presented here allows us to analyze system performance in the presence of large DBCR.

\begin{figure}[htbp]
\centering
\includegraphics{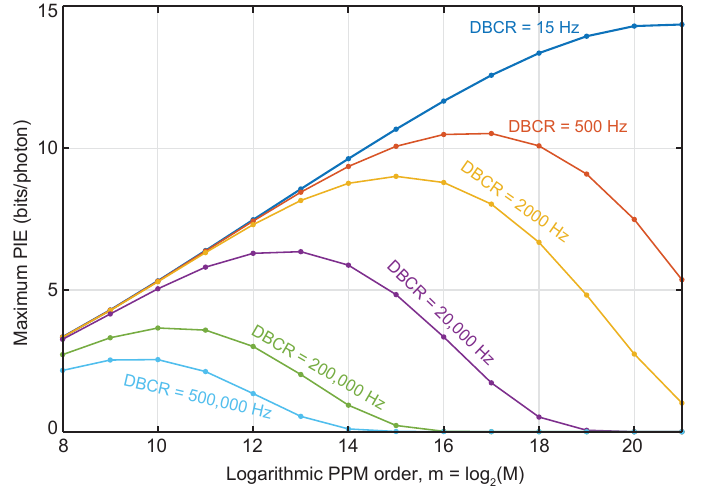}
\caption{Maximum attainable PIE vs logarithmic PPM order $m=\mathrm{log}_2(M)$, computed from theoretical frame erasure and error probabilities, over a range of DBCR values.}
\label{fig:PIEvsMvsDBCR}
\end{figure}

To this end, we use Equations~\ref{eq:PEras} and \ref{eq:PErr} to theoretically compute frame erasure and error ratios over a large range of $\lambda$ (i.e., average optical power incident on the receiver), and calculate the maximum PIE attainable over that range when RS coding is employed. Such a computation is repeated for several values of DBCR. The time slot duration, guard window duration, and SDE chosen for this study are shown in Table~\ref{table:hybridPPMParams}. 

Fig.~\ref{fig:PIEvsMvsDBCR} shows the results. For DBCR = 15 Hz, the maximum attainable PIE begins to saturate close to $m=20$. This results from the fact that the signal count rate corresponding to the $\lambda$ of maximum PIE for $m\sim20$ approaches the DBCR. As the DBCR is increased, the performance of larger $m$ degrades, owing to corruption of their larger-in-number empty timeslots by background radiation, and the most photon efficient order $m$ for a given value of DBCR shifts to a lower value. In other words, in the presence of large DBCR, employing a smaller $m$ is beneficial thanks to their larger frame rate. Put differently, when large background count rates are expected, for example in space missions close to the larger planets or asteroids that are within a distance of $\sim10$~AU from Earth, using a small $m$ is beneficial to maximize data rate (as shown in Fig.~\ref{fig:hybridPPMPerf}) as well as for system performance to be resilient to large background flux (as shown in Fig.~\ref{fig:PIEvsMvsDBCR}).

Notably, a background count rate of $\approx14,000$ Hz was reported in the Pysche mission~\cite{PsycheTechnical2023} at a distance from Earth of $\sim1-3$~AU. For such a DBCR, the most photon-efficient $m$ for the time slot duration chosen here is expected to be $m=13$. Further out in the Solar System, however, background counts may be considerably lower owing to the smaller size of celestial bodies beyond Neptune, as well as the farther distance from the Sun and the Earth. In such a scenario of lower DBCR, employing a large $m$ can help maintain communication over very large distances.

\bibliographystyle{apsrev4-2}
\bibliography{apssamp}
\end{document}